\documentclass[aps,showpacs,floatfix,twocolumn,superscriptaddress,longbibliography]{revtex4-2}

\usepackage{amsmath}
\usepackage{amssymb}
\usepackage{amstext}
\usepackage{amsopn}
\usepackage{amsfonts}
\usepackage{amsxtra}
\usepackage[english]{babel}
\usepackage{graphicx, nicefrac}
\usepackage{float}
\usepackage{siunitx}
\usepackage[latin1]{inputenc}
\usepackage{dcolumn}
\usepackage{bm,graphicx}
\usepackage{etoolbox}
\usepackage[colorlinks]{hyperref}
\usepackage{url}
\usepackage[normalem]{ulem}
\usepackage[dvipsnames,usenames]{color}
\usepackage[version=3]{mhchem} 
\usepackage{booktabs}
\usepackage{tabularx}
\usepackage{hyperref}

\begin{document}


\title{Optical conductivity of layered topological semimetal TaNiTe$_5$}

\author{Jakov Budi\'c}
\altaffiliation[]{These authors contributed equally}
 \affiliation{Department of Physics, Faculty of Science, University of Zagreb, 10000 Zagreb, Croatia}
\author{Serena Nasrallah}
\altaffiliation[]{These authors contributed equally}
\affiliation{Institute of Solid State Physics, TU Wien, 1040 Vienna, Austria}
\author{D. Santos-Cottin}
 \affiliation{Department of Physics, University of Fribourg, 1700 Fribourg, Switzerland}
 \author{F. Le Mardel\'e}
 \affiliation{Department of Physics, University of Fribourg, 1700 Fribourg, Switzerland}
\author{A. Pulkkinen}
\affiliation{New Technologies-Research Centre, University of West Bohemia in Pilsen, 30100 Plze\v{n}, Czech Republic}
\author{J. Min\'ar}
\affiliation{New Technologies-Research Centre, University of West Bohemia in Pilsen, 30100 Plze\v{n}, Czech Republic}
\author{P. Sa\v{c}er}
\affiliation{Department of Physics, Faculty of Science, University of Zagreb, 10000 Zagreb, Croatia}
\author{B. Gudac}
\affiliation{Department of Physics, Faculty of Science, University of Zagreb, 10000 Zagreb, Croatia}
%
%
\author{N. Bari\v{s}i\'c}
\affiliation{Department of Physics, Faculty of Science, University of Zagreb, 10000 Zagreb, Croatia}
\affiliation{Institute of Solid State Physics, TU Wien, 1040 Vienna, Austria}
\author{C.~C. Homes}
\affiliation{National Synchrotron Light Source II, Brookhaven National Laboratory, Upton, New York 11973, USA}
\author{Ana Akrap}
 \email{aakrap.phy@pmf.hr}
\affiliation{Department of Physics, Faculty of Science, University of Zagreb, 10000 Zagreb, Croatia}
\author{Mario Novak}
 \email{mnovak.phy@pmf.hr}
\affiliation{Department of Physics, Faculty of Science, University of Zagreb, 10000 Zagreb, Croatia}

\date{\today}
%
%
\begin{abstract}
We present an infrared spectroscopy study of the layered topological semimetal TaNiTe$_5$, a material with a quasi-one-dimensional structure and strong in-plane anisotropy. Despite its structural features, infrared reflectivity and electronic transport measurements along the $a$ and $c$ crystallographic axes show metallic behavior without evidence of reduced dimensionality. Optical conductivity reveals an anisotropic but conventional metallic response with low scattering rates and a single sharp infrared-active phonon mode at \qty{396}{\per\centi\meter} (\qty{49}{meV}). 
\textit{Ab initio} calculations closely match the experimental optical data and confirm a three-dimensional electronic structure. 
Our results demonstrate that TaNiTe$_5$ behaves as a three-dimensional anisotropic semimetal in its electronic and optical properties.
\end{abstract}

\maketitle

%
%
%
\section{\label{sec:level1} Introduction}
Many emerging technologies increasingly rely on the unique electronic properties of van der Waals telluride
materials \cite{Siddique2021,Banjade2021}. TaNiTe$_5$ is a topological semimetal characterized by a chain-like atomic structure and strong in-plane electronic anisotropy. This material was reported to host a variety of complex electronic structures, including quasi-one-dimensionality, Dirac nodal lines, nodal loops, and fourfold Dirac cones protected
by non-symmorphic symmetry, which remain stable in the presence of spin-orbit coupling \cite{Xu2020,Huang2023,Hao2021, Li2022}. 
TaNiTe$_5$ also shows high magnetoresistance and low effective carrier masses \cite{Ye2022}.
%
Further studies have indicated that TaNiTe$_5$ may exhibit a three-dimensional topological character, supported by 
quantum oscillations \cite{Chen2021} and scanning tunneling microscopy measurements \cite{Ma2023}. This raises questions 
about the actual dimensionality of its electronic structure. To investigate this, we determined the complex optical 
conductivity, which can reveal features of low-dimensional systems, such as van Hove singularities.

%
%
In this work, we have determined the infrared reflectivity and electronic transport along the $a$ and $c$ crystallographic axes. Reflectivity spectra indicate metallic behavior in both directions. Resistivity data show anisotropy and a linear 
temperature dependence above $\approx\qty{50}{\kelvin}$, consistent with a dominant role of electron-phonon scattering. No spectral features indicative 
of reduced dimensionality were observed. The experimental optical response agrees well with \textit{ab initio} calculations.

Our results show that TaNiTe$_5$ is an anisotropic semimetal without clear signatures of low-dimensional electronic 
behavior. The agreement between experiment and \textit{ab initio} calculations further supports a three-dimensional 
electronic structure, despite the quasi-one-dimensional-like material structure.





%
%
\section{\label{sec:level1} Experimental details}
Single crystals of TaNiTe$_5$ have been synthesized using a self-flux method, with the ratio $\ce{Ta}:\ce{Ni}:\ce{Te}=1:1:10$. 
High-purity elements have been sealed in a quartz tube under a vacuum of \qty{e-5}{mbar} and heated to \qty{950}{\celsius}, below the boiling point of tellurium, to dissolve and homogenize the flux. Crystallization was carried out by slow cooling, at a rate of \qty{1}{\kelvin\per\minute}, down to \qty{500}{\celsius}. Single crystals were then extracted by centrifuge. As a result, shiny, easy-to-cleave crystals are obtained with lateral dimensions ($a$-$c$ plane) of about $\qty{5}{\milli\meter}\times\qty{2}{\milli\meter}$. 
The crystals were characterized by powder X-ray diffraction (PXRD), energy-dispersive X-ray spectroscopy (EDX), and Laue diffraction to confirm their crystal structure, composition, and orientation, respectively.

Resistivity and Hall effect were measured using a home-built setup using the four-contact method.
Infrared reflectance was measured using a Bruker Vertex 70v and 80v Fourier transform spectrometers. The reflectivity measurement is performed at a near-normal angle of incidence. Temperature control was achieved using a continuous-flow cryostat with a cold finger.
In order to measure the absolute reflectance, an {\em in situ} gold evaporation technique was used \cite{Homes1993}. 
After the spectra of the sample were measured, a tungsten wire coil was used to evaporate a small amount of gold onto the sample. 
The measurements were then repeated on the now gold-coated sample. This method allows us to have an almost perfect reference because it takes into account the shape, surface irregularities, and position of the sample.

%
%
The reflectivity is a combination of the real and imaginary parts of the dielectric function, and as such can be difficult to interpret. The real part of the optical conductivity, calculated from the imaginary part of the dielectric function, is a more intuitive quantity. Accordingly, the complex dielectric function, $\tilde\epsilon(\omega) = \epsilon_1 + i\epsilon_2$, has been determined from a Kramers-Kronig analysis of the reflectivity, which requires extrapolations at high and low frequencies. At low frequency a metallic Hagen-Rubens extrapolation, $R(\omega) = 1 - A\sqrt{\omega}$ was employed, where $A$ is chosen to match the value of the reflectance at the lowest measured frequency.  At high frequency, the reflectivity measurements were extended using ellipsometry results up to 5~eV. At energies above that, the reflectivity can be extrapolated from X-ray scattering functions (XRO) \cite{Tanner2015, Kuzmenko2005}. 
The complex conductivity, $\tilde\sigma(\omega)$, is calculated from the complex dielectric function,
$\tilde\sigma(\omega) = \sigma_1 +i\sigma_2 = i \frac{2\pi}{Z_0} \omega [1 - \tilde\epsilon(\omega)]$, 
where $Z_0 \approx \qty{377}{\ohm}$ is the impedance of free space and $\omega$ is in units of $\qty{}{\per\centi\meter}$.
%
%

\textit{Ab initio} electronic structure calculations were performed with the linearized augmented wave method with 
local orbitals (LAPW+lo) using the Elk software \cite{elk}. The basis set cutoff parameter $R_{\text{MT}} K_{\text{max}}$ 
was set to \qty{8.0}{}, and the exchange-correlation effects were included at the level of the generalized gradient approximation 
(GGA) as implemented in the PBE functional \cite{PhysRevLett.77.3865}. The optical conductivity was calculated using a 
dense sampling of the Brillouin zone with a $45\times 45 \times 21$ $k$-point mesh. The occupation numbers were smeared 
using the Fermi-Dirac distribution with a smearing width of \qty{0.2}{mHa}. Spin-orbit interaction was included in all 
calculations. The calculated optical conductivities were broadened by convolution with a Lorentzian function with full 
width at half maximum \qty{50}{meV}, and the Drude term describing intraband transitions was added after broadening.

%
%
\section{\label{sec:level1} Results and Discussion}
\begin{figure}[t]
    \centering
    \includegraphics[width=0.9\linewidth]{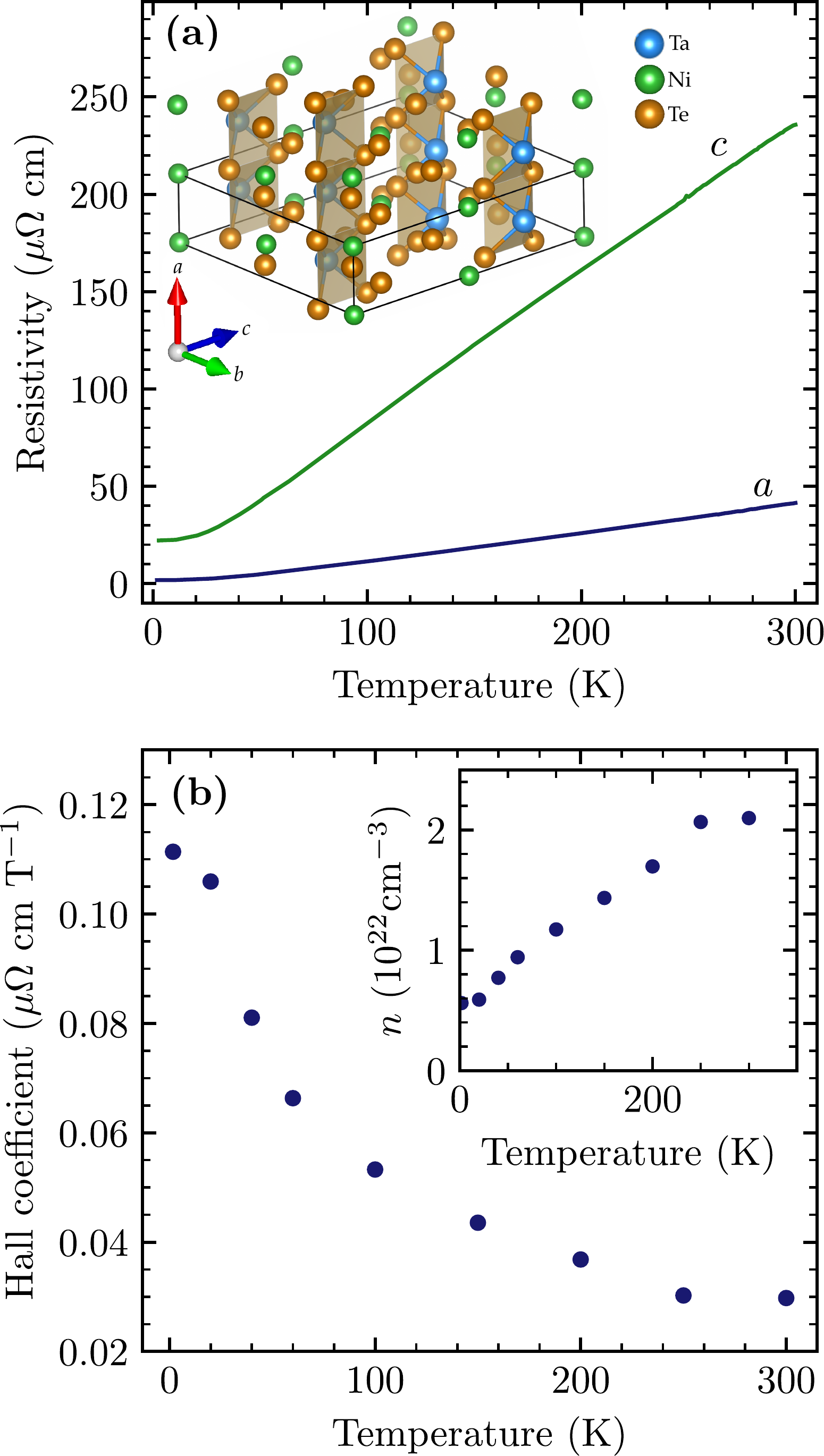}
    \caption{(a) The experimentally-determined temperature dependence of resistivity ($\rho$) for TaNiTe$_5$ along $a$ and $c$ axes. 
    The inset shows \ce{TaNiTe5} structure \cite{VESTA}. (b) Measured temperature dependence of Hall coefficient $R_H$ for the same 
    sample. The inset shows carrier density at different temperatures calculated using a simple model $n\propto 1/R_H$.}
    \label{fig1}
\end{figure}

The layered TaNiTe$_5$ material belongs to the class of van der Waals materials, with an easy cleaving $a-c$ plane perpendicular 
to the crystallographic $b$ axis. It crystallizes in an orthorhombic crystal structure Cmcm (No. 63) with lattice parameters 
$a=3.659$~\AA , $b=13.122$~\AA , and $c=15.111$~\AA\ \cite{Liimatta1989}. The structure is shown in the inset of Fig.~\ref{fig1}(a). TaNiTe$_5$ is composed of 1D polymeric chains of nickel and tantalum along the $a$ direction, bridged by tellurium atoms. Nickel atoms are octahedrally coordinated, and tantalum atoms are in a strongly disordered square-pyramidal coordination with ligand tellurium atoms.

The resistivity of TaNiTe$_5$ is shown in Fig.~\ref{fig1} (a). It is metallic in character along both the $a$ and $c$ directions, 
with a linear temperature dependence above $\approx\qty{50}{K}$, consistent with electron-phonon scattering.
We observe a noticeable anisotropy: $\rho_a/\rho_c\approx\qty{0.08}{}$ at $\qty{1.8}{\kelvin}$ and $\rho_a/\rho_c\approx\qty{0.18}{}$ 
at $\qty{300}{\kelvin}$. The residual resistivity is much larger in the $c$-direction, and the temperature dependence is stronger.
The Hall coefficient $R_H$ does not change sign in the measured temperature range, from $\qty{1.8}{\K}$ up to $\qty{300}{K}$. 
Assuming a single band, $R_H = 1/ne$, the number of carriers is of the order of magnitude $n\approx \qty{E22}{\per\centi\meter\cubed}$, 
consistent with the metallic resistivity. The measured $R_H(T)$ implies that $n$ decreases at low temperatures, consistent with a 
semimetallic band structure.
%
%
%

\begin{figure*}[ht]
    \centering
    \includegraphics[width=0.9\linewidth]{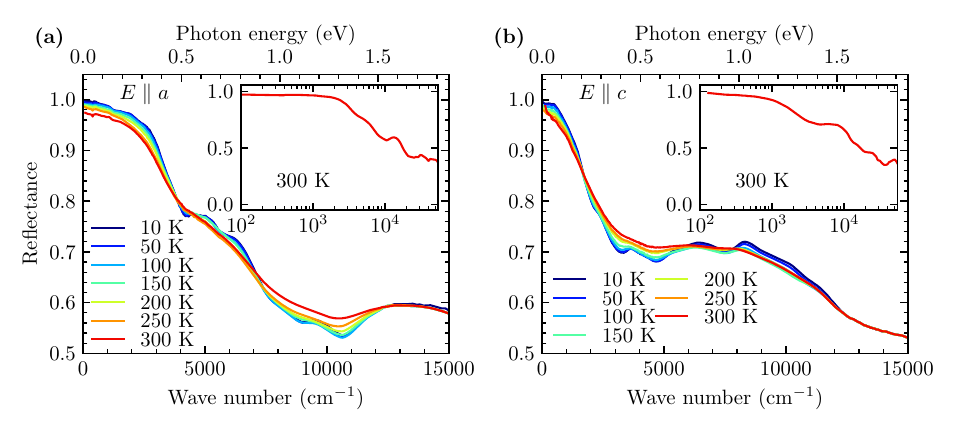}
    \caption{The temperature dependence of the reflectivity of a TaNiTe$_5$ single crystal as a function of wave number (photon energy) 
      along the (a) $a$ axis, and (b) $c$ axis, up to $\qty{15000}{\per\centi\meter}$, above which the temperature dependence is 
      insignificant. The insets show the full range of measured reflectance at room temperature.}
    \label{fig2}
\end{figure*}
The temperature dependence of the reflectivity measured over a wide frequency range for light polarized along the $a$ and $c$ axes is shown in Figs.~\ref{fig2}(a) and \ref{fig2}(b), respectively. Despite a large anisotropy in the reflectivity, the response is metallic in both cases, with $R \rightarrow 1$ in the $\hbar \omega \rightarrow 0$ limit. The temperature dependence was determined up to $\qty{15000}{\per\centi\meter}$, above which it is assumed to have little temperature dependence. The full range, up to $\qty{40000}{\per\centi\meter}$, at room temperature, is shown in the insets.

\begin{figure*}[t]
    \centering    \includegraphics[width=\linewidth]{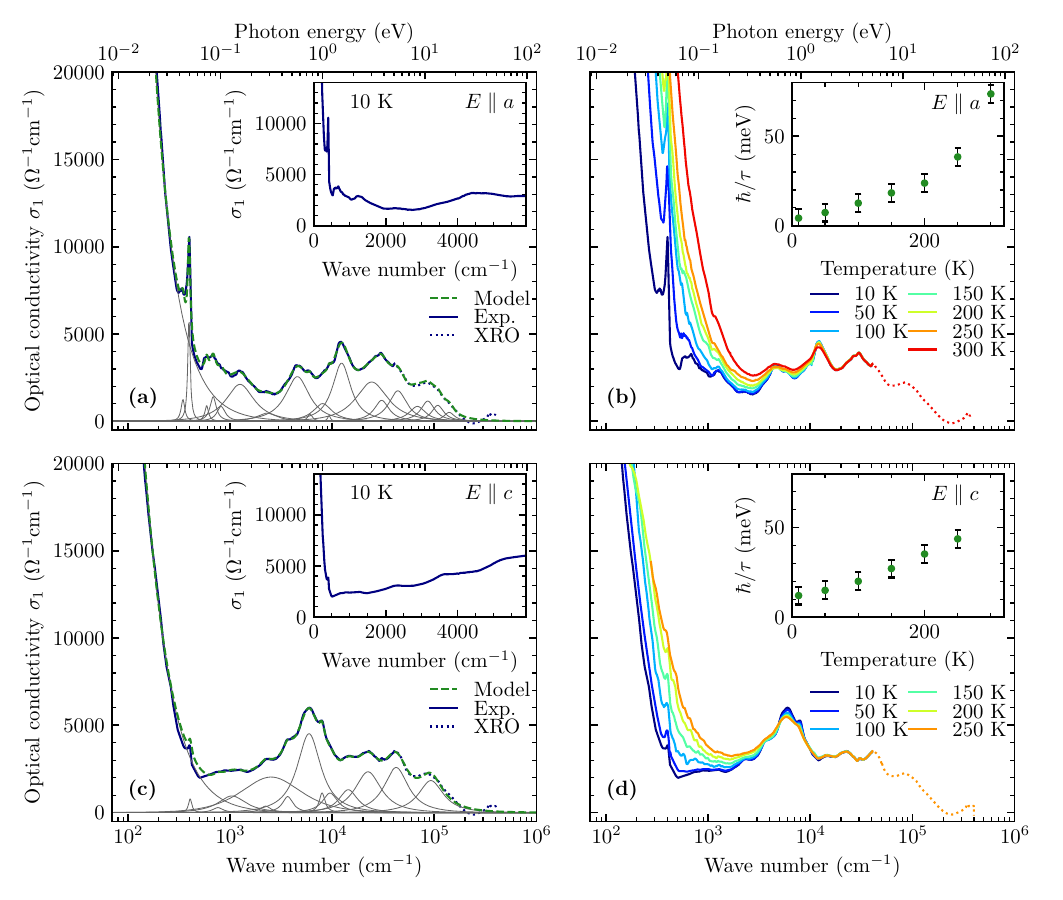}
    \caption{(a) The the real part of the optical conductivity ($\sigma_1$) of TaNiTe$_5$ along the $a$ axis at 10~K.  
    The conductivity decomposition was obtained by fitting $\sigma_1(\omega)$ to the Drude-Lorentz sum, Eq.~(\ref{eq:dl}).
    The Drude-Lorentz conductivity fit is shown by the dashed line. It contains a strong and narrow Drude response, characterized as Lorentzian centered as zero frequency with a full width at half maximum of $1/\tau$, 
    a sharp and narrow lattice contribution (phonon line), and a number of bound excitations attributed to interband transitions. 
    The dotted line shows data that was extrapolated from X-ray scattering functions (XRO) \cite{Tanner2015}. The inset shows the 
    low-energy range of the optical conductivity with a linear scale. (b) The temperature dependence of the real part of optical 
    conductivity along the $a$ direction.  The inset shows temperature dependence of scattering rate acquired from Drude-Lorentz 
    fits. Dotted line shows data that was generated from XRO. 
    This analysis has been repeated for the $c$-axis optical conductivity, shown in (c) and (d).}
    \label{fig3}
\end{figure*}

To further analyze the optical data, we use Kramers-Kronig transformations to obtain complex optical response functions, specifically
the optical conductivity and the dielectric function. The real part of the optical conductivity $\sigma_1$ at $\qty{10}{\K}$ is shown in Figs.~\ref{fig3}(a) and \ref{fig3}(c) for light polarized along the $a$ and $c$ axis, respectively. The insets show $\sigma_1$ in a lower energy range on a linear scale. Note that the optical data at energies higher than \qty{40000}{\per\centi\meter} is extrapolated from X-ray scattering functions (XRO) \cite{Tanner2015}.

%
%

%

%

To investigate the behavior of the optical properties in more detail, the optical response has been modeled using the Drude-Lorentz model for the complex dielectric function
%
%
\begin{equation}
  \tilde\epsilon(\omega) = \epsilon_\infty - \frac{\omega_{p}^2}{\omega^2+i\omega/\tau}
    + \sum_j \frac{\Omega_j^2}{\omega_j^2 - \omega^2 - i\omega\gamma_j}.
  \label{eq:dl}
\end{equation}
In the first term $\omega_{p}^2 = 4\pi ne^2/m^\ast$ and $1/\tau$ are
the square of the plasma frequency and scattering rate for the delocalized (Drude)
carriers, respectively, and $n$ and $m^\ast$ are the carrier concentration and effective mass.
In the summation, $\omega_j$, $\gamma_j$ and $\Omega_j$ are the position, width, and
strength of a symmetric Lorentzian oscillator that describe the $j$th vibration or bound
excitation.  The term $\epsilon_\infty$ is meant to capture the contributions to the real part of the dielectric function from high-frequency excitations.

%

The infrared reflectance was fitted using a nonlinear least-squares approach, and then the optical conductivity was calculated.  In addition to the bound excitations, the anisotropic response of the free carriers is revealed through the plasma frequency and the scattering rate; $1/\tau$ is shown in the insets of Figs.~\ref{fig3}(b) and \ref{fig3}(d) as a function of temperature along the $a$ and $c$ axes, respectively. The plasma frequency calculated from the Drude-Lorentz fit is roughly temperature independent and amounts to $\omega_{p,a} \approx \qty{32000}{\per\centi\meter}$ in $a$ direction and $\omega_{p,c}\approx\qty{21000}{\per\centi\meter}$ in $c$ direction.
%
%
The ability to separate the two Drude parameters, $1/\tau$ and $\omega_p$, is an advantage of optical conductivity. The Drude plasma frequency $\omega_p$, and the scattering rate can be related to resistivity, 
\begin{equation}
    \rho = \frac{Z_0}{2\pi} \frac{1}{\omega_p^2 \tau},
\end{equation}
where $Z_0 \approx \qty{377}{\ohm}$ is the impedance of free space, and $\omega_p$ is in units of $\qty{}{\per\centi\meter}$.
However, the low-frequency reflectance in \ce{TaNiTe5} is quite close to unity. Since $\sigma_1(\omega) \propto 1/[1-R(\omega)]$, a small uncertainty in reflectance in this region significantly affects optical conductivity. This yields a relatively high uncertainty for the scattering rate $1/\tau$.

%
The scattering rate $1/\tau$ at lower temperatures is higher for the $c$ direction, while the plasma frequency $\omega_p$ has little to no temperature dependence. The results are therefore consistent with the resistivity data.
The temperature dependence of the scattering rate seems to be linear, especially at lower temperatures. 
Furthermore, the anisotropy can be seen between $a$ and $c$ directions; however, it becomes less apparent as the temperature increases.

A single infrared active phonon can be seen at $\approx\qty{396}{\per\centi\meter}$, in both directions, although much sharper when light was polarized parallel to the $a$-axis.
The irreducible vibrational representation of \ce{TaNiTe5} is: 
$$
\Gamma_{\rm irr} = 10 A_g + 10 B_{1g} + 8 B_{2g} + 8 B_{3g} + 9 B_{1u} + 10 B_{2u} + 11 B_{3u}.
$$
The $A_g$ and $B_g$ modes are Raman active, while only the $B_u$ modes are infrared active. A calculation using an 
empirical model normal coordinate analysis (force field model) \cite{vibratz} gives one strong mode along the $a$ axis, 
which we identify as a $B_{3u}$ mode, at a frequency consistent with our measurements.

The optical response at energies above the Drude term contains multiple interband transitions which we fit by several Lorentzian oscillators in the Drude-Lorentz model, shown in Figs. \ref{fig3}(a) and \ref{fig3}(c). There are no apparent traces of low dimensionality in $\sigma_1(\omega)$ in $a$ and $c$ directions, contrary to what is expected for a quasi-1D material \cite{Homes2018}. A low-dimensional band dispersion leads to a flat band for certain $k$-space directions. This results in van Hove singularities \cite{CardonaYu}. In optical conductivity, such van Hove singularities would be visible as sharp peaks or sharp onsets of absorption.

%
%
\begin{figure*}[!t]
    \centering    
    \includegraphics[width=0.9\linewidth]{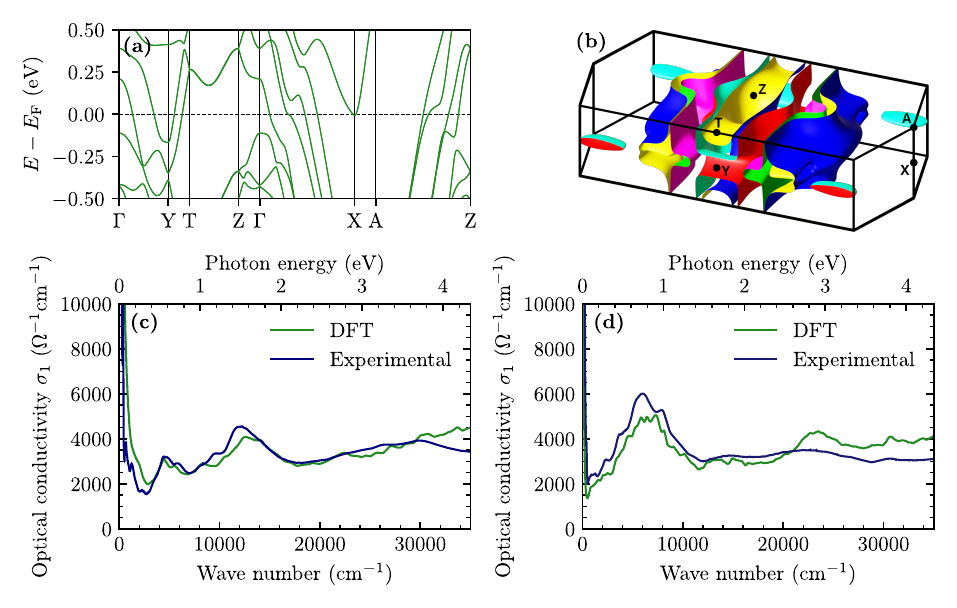}
    \caption{(a) Calculated band structure for TaNiTe$_5$ along the high-symmetry directions. (b) Calculated Fermi surface, which 
    contains strongly warped 2D sheets and 3D pockets.  (c) Comparison between the DFT-calculated and experimental real parts of the optical 
    conductivity along the $a$ axis, and (d) along the $c$ axis.}
    \label{fig4}
\end{figure*}

Since \ce{TaNiTe5} has been reported to be a nodal line semimetal, it makes sense to compare it with other such materials, 
for example the representative square net lattice systems \ce{ZrSiS} \cite{Schilling2017}, \ce{ZrSiSe} \cite{Shao2020} and 
\ce{BaNiS2} \cite{Santos-Cottin2021}. In all of these nodal line semimetals, there is a fairly flat part in $\sigma_1(\omega)$, 
with little frequency dependence, and a temperature-induced spectral weight transfer. The spectral weight moves from energies 
above the region of flat $\sigma_1(\omega)$ to energies below that region. No similar behavior can be seen in \ce{TaNiTe5}. Instead, a strong Drude contribution is followed by a series of broad, frequency-dependent excitations.

Figure~\ref{fig4}(a) shows the calculated band structure along the high-symmetry directions, resulting in a composite Fermi surface shown in Fig.~\ref{fig4}(b). 
As a result of structural features, the calculated Fermi surface consists of strongly warped 2D sheets, an interconnected tubular pocket with open orbits along both the $c$ and $b$ directions, and ellipsoidal 3D pockets.
The 2D sheets reflect a structural motif of tantalum and nickel polymer chains running along the $a$ direction.
From this calculation, the optical conductivity, shown in Figs.~\ref{fig4}(c) and \ref{fig4}(d) 
for the two light polarizations, can be determined. Since the band structure shows multiple bands crossing the Fermi level, we expect multiple Drude contributions in optical conductivity. 
It is worth noting that while multiple Drude components were introduced into the fits, in each case they collapsed back into a single Drude term, suggesting that while this is clearly a multiband material, the scattering rates appear to be similar enough so that a single Drude term describes the real part of the optical conductivity.
Furthermore, multiple bands above and below the Fermi level allow for a number of interband transitions to be excited. Many of these bands are parallel, leading to an enhanced joint density of states. Each of the transitions can be seen as a feature in the optical conductivity, 
providing an explanation for multiple peaks in the measured data.
In general, the \textit{ab initio} calculations correspond very well to the experimental optical conductivity. 

%
%
\section{\label{sec:level1} Conclusion}
We have measured the temperature dependence of the electronic transport and determined the complex optical properties of 
\ce{TaNiTe5} along the $a$ and $c$ crystallographic axes. The optical conductivity shows metallic-like behavior in both 
directions. Resistivity measurements show noticeable anisotropy and linear temperature dependence above $\approx\qty{50}{\kelvin}$ consistent 
with electron-phonon scattering.  Using Kramers-Kronig transformations, we calculated optical response functions and 
analyzed them using the Drude-Lorentz model. {\em Ab initio} calculations agree very well with the optical conductivity 
data. This gives confidence that the calculated band structure of \ce{TaNiTe5} correctly describes this material.
Infrared spectroscopy shows that \ce{TaNiTe5} is an anisotropic semimetal with low scattering rates and pronounced interband 
transitions. Neither in the calculation nor in the infrared measurements can one see any clear signatures of reduced dimensionality.


\section{\label{sec:level1} Acknowledgments}

This research was supported by the NCCR MARVEL, a National Centre of Competence in Research, funded by the Swiss National Science Foundation (Grant No. 205602).  A.A. acknowledges funding from the Swiss National Science Foundation through Project No. PP00P2\_202661. Work at Brookhaven National Laboratory was supported by the Office of Science, U.S. Department of Energy, under Contract No. DE-SC0012704. N.B., A.A. and J.B. acknowledge the support of the Croatian Science Foundation under Project No. IP-2022-10-3382. M.N. and N.B. acknowledge the support of the CeNIKS project co-financed by the Croatian Government and the EU through the European Regional Development Fund Competitiveness and Cohesion Operational Program (Grant No. KK.01.1.1.02.0013). A.A. and M.N. acknowledge funding from the Swiss National Science Foundation, through the instrument Scientific Exchanges, Grant number 199151. J.M. and A.P. were supported by the project Quantum
materials for applications in sustainable technologies (QM4ST), funded
as project No. CZ.02.01.01/00/22\_008/0004572 by P JAK, call Excellent
Research.
The work at the TU Wien was supported by FWF Project P 35945-N.

\textit{Data availability.} The data that support the findings of this article are openly available \cite{data}.


%
%
%

\bibliography{references}

@article{Banjade2021,
  title = {Monolayer 2D semiconducting tellurides for high-mobility electronics},
  author = {Banjade, Huta R. and Pan, Jinbo and Yan, Qimin},
  journal = {Phys. Rev. Mater.},
  volume = {5},
  issue = {1},
  pages = {014005},
  numpages = {6},
  year = {2021},
  month = {Jan},
  publisher = {American Physical Society},
  doi = {10.1103/PhysRevMaterials.5.014005},
}

@book{CardonaYu,
	address = {Berlin Heidelberg},
	author = {P. Y. Yu and M. Cardona},
	isbn = {978-3-642-00709-5},
	publisher = {Springer},
	title = {Fundamentals of Semiconductors},
	year = {2010}
}

@article{Chen2021,
	author = {Chen, Zheng and Wu, Min and Zhang, Yong and Zhang, Jinglei and Nie, Yong and Qin, Yongliang and Han, Yuyan and
      Xi, Chuanying and Ma, Shuaiqi and Kan, Xucai and Zhou, Jianhui and Yang, Xiaoping and Zhu, Xiangde and Ning, Wei and Tian, Mingliang},
	doi = {10.1103/PhysRevB.103.035105},
	journal = {Phys. Rev. B},
	month = {01},
	number = {3},
	pages = {035105},
	publisher = {American Physical Society},
	title = {{Three-dimensional topological semimetal phase in layered TaNiTe$_5$ probed by quantum oscillations}},
	volume = {103},
	year = {2021},
}

@misc{elk,
	howpublished = {\url{http://elk.sourceforge.net/}},
	title = {{The Elk Code}}
}

@article{Hao2021,
	author = {Hao, Zhanyang and Chen, Weizhao and Wang, Yuan and Li, Jiayu and Ma, Xiao-Ming and Hao, Yu-Jie and Lu, Ruie and
      Shen, Zecheng and Jiang, Zhicheng and Liu, Wanling and Jiang, Qi and Yang, Yichen and Lei, Xiao and Wang, Le and Fu, Ying and
      Zhou, Liang and Huang, Lianglong and Liu, Zhengtai and Ye, Mao and Shen, Dawei and Mei, Jiawei and He, Hongtao and Liu, Cai and
      Deng, Ke and Liu, Chang and Liu, Qihang and Chen, Chaoyu},
	doi = {10.1103/PhysRevB.104.115158},
	journal = {Phys. Rev. B},
	number = {11},
	pages = {115158},
	publisher = {American Physical Society},
	title = {{Multiple Dirac nodal lines in an in-plane anisotropic semimetal TaNiTe$_5$}},
	volume = {104},
	year = {2021},
}

@article{Homes1993,
	author = {Homes, Christopher C. and Reedyk, M. and Crandles, D. A. and Timusk, T.},
	doi = {10.1364/AO.32.002976},
	journal = {Appl. Opt.},
	number = {16},
	pages = {2976--2983},
	publisher = {OSA},
	title = {Technique for measuring the reflectance of irregular, submillimeter-sized samples},
	volume = {32},
	year = {1993},
}

@article{Homes2018,
	author = {Homes, C. C. and Du, Q. and Petrovic, C. and Brito, W. H. and Choi, S. and Kotliar, G.},
	doi = {10.1038/s41598-018-29909-2},
	isbn = {2045-2322},
	journal = {Scientific Reports},
	number = {1},
	pages = {11692},
	title = {{Unusual electronic and vibrational properties in the colossal thermopower material FeSb$_2$}},
	volume = {8},
	year = {2018},
}

@article{Huang2023,
	title = {{Magnetic Field-Induced Resistivity Upturn and Non-Topological Origin in the Quasi-One-Dimensional Semimetals}},
	author = {Huang, Yalei and Ye, Rongli and Shen, Weihao and Yao, Xinyu and Cao, Guixin},
	journal = {Symmetry},
	number = {10},
	volume = {15},
    pages = {1882},
	year = {2023},
	doi = {10.3390/sym15101882},
}

@article{Kuzmenko2005,
	author = {A. B. Kuzmenko},
	doi = {10.1063/1.1979470},
	journal = {Review of Scientific Instruments},
	number = {8},
	numpages = {9},
	pages = {083108},
	publisher = {AIP},
	title = {{Kramers--Kronig constrained variational analysis of optical spectra}},
	volume = {76},
	year = {2005},
}

@article{Li2022,
  title = {{Coexistence of Ferroelectriclike Polarization and Dirac-like Surface State in TaNiTe$_5$}},
  author = {Li, Yunlong and Ran, Zhao and Huang, Chaozhi and Wang, Guanyong and Shen, Peiyue and Huang, Haili and Xu, Chunqiang and
    Liu, Yi and Jiao, Wenhe and Jiang, Wenxiang and Hu, Jiayuan and Zhu, Gucheng and Xu, Chenhang and Lu, Qi and Wang, Guohua and
    Jing, Qiang and Wang, Shiyong and Shi, Zhiwen and Jia, Jinfeng and Xu, Xiaofeng and Zhang, Wentao and Luo, Weidong and Qian, Dong},
  journal = {Phys. Rev. Lett.},
  volume = {128},
  issue = {10},
  pages = {106802},
  numpages = {6},
  year = {2022},
  month = {Mar},
  publisher = {American Physical Society},
  doi = {10.1103/PhysRevLett.128.106802},
}

@article{Liimatta1989,
	author = {Liimatta, Eric W. and Ibers, James A.},
	doi = {https://doi.org/10.1016/0022-4596(89)90122-9},
	isbn = {0022-4596},
	journal = {Journal of Solid State Chemistry},
	number = {1},
	pages = {7--16},
	title = {{Synthesis, structures, and conductivities of the new layered compounds Ta$_3$Pd$_3$Te$_{14}$ and TaNiTe$_5$}},
	url = {https://www.sciencedirect.com/science/article/pii/0022459689901229},
	volume = {78},
	year = {1989},
}

@article{Ma2023,
	author = {Ma, Ni and Wang, De-Yang and Huang, Ben-Rui and Li, Kai-Yi and Song, Jing-Peng and Liu, Jian-Zhong and
      Mei, Hong-Ping and Ye, Mao and Li, Ang},
	doi = {10.1088/1674-1056/aca203},
	journal = {Chinese Physics B},
	number = {5},
	pages = {056801},
	publisher = {Chinese Physical Society and IOP Publishing Ltd},
	title = {{Quasi-one-dimensional characters in topological semimetal TaNiTe$_5$}},
	volume = {32},
	year = {2023},
}

@article{PhysRevLett.77.3865,
	author = {Perdew, John P. and Burke, Kieron and Ernzerhof, Matthias},
	doi = {10.1103/PhysRevLett.77.3865},
	issue = {18},
	journal = {Phys. Rev. Lett.},
	month = {Oct},
	pages = {3865--3868},
	publisher = {American Physical Society},
	title = {{Generalized Gradient Approximation Made Simple}},
	volume = {77},
	year = {1996},
}

@article{Santos-Cottin2021,
  title = {{Optical conductivity signatures of open Dirac nodal lines}},
  author = {Santos-Cottin, David and Casula, Michele and de' Medici, Luca and Le Mardel{\'e}, F. and Wyzula, J. and Orlita, M. and
    Klein, Yannick and Gauzzi, Andrea and Akrap, Ana and Lobo, R. P. S. M.},
  journal = {Phys. Rev. B},
  volume = {104},
  issue = {20},
  pages = {L201115},
  numpages = {6},
  year = {2021},
  month = {Nov},
  publisher = {American Physical Society},
  doi = {10.1103/PhysRevB.104.L201115},
}

@article{Schilling2017,
	author = {Schilling, M. B. and Schoop, L. M. and Lotsch, B. V. and Dressel, M. and Pronin, A. V.},
	doi = {10.1103/PhysRevLett.119.187401},
	journal = {Phys. Rev. Lett.},
	month = {11},
	number = {18},
	pages = {187401},
	publisher = {American Physical Society},
	title = {{Flat Optical Conductivity in ZrSiS due to Two-Dimensional Dirac Bands}},
	volume = {119},
	year = {2017},
}

@article{Shao2020,
	author = {Shao, Yinming and Rudenko, A. N. and Hu, Jin and Sun, Zhiyuan and Zhu, Yanglin and Moon, Seongphill and
      Millis, A. J. and Yuan, Shengjun and Lichtenstein, A. I. and Smirnov, Dmitry and Mao, Z. Q. and Katsnelson, M. I. and Basov, D. N.},
	doi = {10.1038/s41567-020-0859-z},
	journal = {Nature Physics},
	number = {6},
	pages = {636--641},
	title = {Electronic correlations in nodal-line semimetals},
	volume = {16},
	year = {2020},
}

@article{Siddique2021,
  title   = {Emerging two-dimensional tellurides},
  author  = {Saif Siddique and Chinmayee Chowde Gowda and Solomon Demiss and Raphael Tromer and Sourav Paul and Kishor Kumar and
    Kishor Kumar Sadasivuni and Emmanuel Femi Olu and Amreesh Chandra and Vidya Kochat and Douglas S. Galv\~{a}o and Partha Kumbhakar and
    Rohan Mishra and Pulickel M. Ajayan and Chandra Sekhar Tiwary},
  journal = {Materials Today},
  pages   = {402--426},
  volume  = {51},
  month   = {December},
  year    = {2021},
  doi     = {10.1016/j.mattod.2021.08.008},
}

@article{Tanner2015,
	author = {Tanner, D. B.},
	doi = {10.1103/PhysRevB.91.035123},
	journal = {Phys. Rev. B},
	month = {01},
	number = {3},
	pages = {035123},
	publisher = {American Physical Society},
	title = {{Use of x-ray scattering functions in Kramers-Kronig analysis of reflectance}},
	volume = {91},
	year = {2015},
}

@article{VESTA,
  author  = {K. Momma and F. Izumi},
  title   = {{\it VESTA 3} for three-dimensional visualization of crystal, volumetric and morphology data},
  journal = {J. Appl. Crystr.},
  volume  = {44},
  number  = {6},
  pages   = {1272-1276},
  month   = {Dec},
  year    = {2011},
  doi     = {10.1107/S0021889811038970},
  URL     = {https://doi.org/10.1107/S0021889811038970},
}

@misc{vibratz,
	note = {E. Dowty, computer code, VIBRATZ (Shape Software, Kingsport, TN, 2001)}
}

@article{Xu2020,
	author = {Xu, Chunqiang and Liu, Yi and Cai, Pinggen and Li, Bin and Jiao, Wenhe and Li, Yunlong and Zhang, Junyi and
      Zhou, Wei and Qian, Bin and Jiang, Xuefan and Shi, Zhixiang and Sankar, Raman and Zhang, Jinglei and Yang, Feng and
      Zhu, Zengwei and Biswas, Pabitra and Qian, Dong and Ke, Xianglin and Xu, Xiaofeng},
	doi = {10.1021/acs.jpclett.0c02382},
	journal = {J. Phys. Chem. Lett.},
	month = {Sep},
	number = {18},
	pages = {7782--7789},
	title = {{Anisotropic Transport and Quantum Oscillations in the Quasi-One-Dimensional TaNiTe$_5$: Evidence for the Nontrivial Band Topology}},
	volume = {11},
	year = {2020},
}

@article{Ye2022,
	author = {Ye, Rongli and Gao, Tian and Li, Haoyu and Liang, Xiao and Cao, Guixin},
	doi = {10.1063/5.0086414},
	isbn = {2158-3226},
	journal = {AIP Advances},
	number = {4},
	pages = {045104},
	title = {{Anisotropic giant magnetoresistanceand de Hass--van Alphen oscillations in layered topological semimetal crystals}},
	volume = {12},
	year = {2022},
}

@dataset{data,
  author       = {Budic, Jakov},
  title        = {Optical conductivity of layered topological
                   semimetal {TaNiTe$_5$}
                  },
  month        = nov,
  year         = 2025,
  publisher    = {Zenodo},
  doi          = {10.5281/zenodo.17601610},
  url          = {https://doi.org/10.5281/zenodo.17601610},
}

\end{document}